\documentclass[fleqn,twoside]{article}
\usepackage{espcrc2}

\usepackage{graphicx}
\usepackage[figuresright]{rotating}

% ...

\def\ARAA{{\it Annual Rev. of Astron. \& Astrophys.} }
\def\ApJ{{\it Astrophys. J.} }
\def\ApJL{{\it Astrophys. J. Letters} }
\def\ApJS{{\it Astrophys. J. Suppl.} }
\def\ApP{{\it Astropart. Phys.} }
\def\AA{{\it Astron. \& Astroph.} }

\def\AAL{{\it Astron. \& Astroph. Letters} }
\def\AASu{{\it Astron. \& Astroph. Suppl.} }

\def\MNRAS{{\it Month. Not. Roy. Astr. Soc.} }
\def\Nature{{\it Nature} }
\def\NewAR{{\it New Astron. Rev.} }

\def\PR{{\it Phys. Rev.} }
\def\PRD{{\it Phys. Rev.} {\bf D} }
\def\PRL{{\it Phys. Rev. Letters} }
\def\RMP{{\it Rev. Mod. Phys.} }
\def\Science{{\it Science} }

\def\etal{{\it et al.}}
\def\simle{\lower 2pt \hbox {$\buildrel < \over {\scriptstyle \sim }$}}
\def\simge{\lower 2pt \hbox {$\buildrel > \over {\scriptstyle \sim }$}}

\newcommand{\AmS}{{\protect\the\textfont2
A\kern-.1667em\lower.5ex\hbox{M}\kern-.125emS}}

\title{Active Galactic Nuclei: Sources for ultra high energy cosmic rays?}

\author{Peter L. Biermann\address[MPI]{MPI for Radioastronomy, Bonn, 
Germany}
\address{Dept. of Phys. \& Astron., Univ. of Bonn, Germany}
\address{Dept. of Phys. \& Astr., Univ. of Alabama, Tuscaloosa, AL, USA}
\address{Dept. of Phys., Univ. of Alabama at Huntsville, AL, USA}
\address[KIT]{Inst. Nucl. Phys. FZ, Karlsruhe Inst. of Techn. (KIT), 
Germany},
Julia K. Becker\address{Institution f{\"o}r Fysik, G{\"o}teborgs Univ., 
Sweden}
\address{Dept. of Phys., Univ. Dortmund, Dortmund, Germany},
Lauren\c{t}iu Caramete\addressmark[MPI]
\address[ROM]{Institute for Space Studies, Bucharest, Romania},
Alex Curu\c{t}iu\addressmark[MPI],
Ralph Engel\addressmark[KIT],
Heino Falcke\address{Dept. of Astrophys., IMAP, Radboud Univ., Nijmegen, 
Netherlands}
\address{ASTRON, Dwingeloo, Netherlands},
L\'aszl\'o {\'A.} Gergely\address{Dept. Appl. Sci., London South Bank 
University, UK}
\address{Dept. of Theoret. \& Exp. Phys., Univ. of Szeged, Szeged, Hungary},
P. Gina Isar\addressmark[KIT]
\addressmark[ROM],
Ioana C. Mari\c{s}\addressmark[KIT],
Athina Meli\address{Physik. Inst. Univ. Erlangen-N{\"u}rnberg, Germany},
Karl-Heinz Kampert\address[WUP]{Phys. Dept., Univ. Wuppertal, Germany},
Todor Stanev\address{Bartol Research Inst., Univ. of Delaware, Newark, 
DE, USA},
Oana Ta\c{s}c\u{a}u\addressmark[WUP],
Christian Zier\addressmark[MPI]
\address{Raman Res. Inst., Bangalore, India}}

\begin{document}

\begin{abstract}
The origin of ultra high energy cosmic rays promises to lead us to a 
deeper understanding of the structure of matter. This is possible 
through the study of particle collisions at center-of-mass energies in 
interactions far larger than anything possible with the Large Hadron 
Collider, albeit at the substantial cost of no control over the sources 
and interaction sites. For the extreme energies we have to identify and 
understand the sources first, before trying to use them as physics 
laboratories. Here we describe the current stage of this exploration. 
The most promising contenders as sources are radio galaxies and gamma 
ray bursts. The sky distribution of observed events yields a hint 
favoring radio galaxies. Key in this quest are the intergalactic and 
galactic magnetic fields, whose strength and structure are not yet fully 
understood. Current data and statistics do not yet allow a final 
judgement. We outline how we may progress in the near future.
\vspace{1pc}
\end{abstract}

\maketitle

\section{Introduction}

In the quest of understanding the fundamental structure of matter the 
physics community has built the Large Hadron Collider (LHC) at CERN in 
Geneva, which will have particle energies up into the TeV region. In the 
universe, as we know from direct observation (first Linsley 1963), we 
have particles up to 300 EeV (=$\; 3 \cdot 10^{20}$ eV); even in the 
center of mass collision with an identical particle (proton or heavier) 
this yields energies in the center of mass frame of $5 \cdot 10^{14}$ 
eV, so more. If we could identify the sources and interaction regions of 
these extreme energy particles (see the books by Ginzburg \& Syrovatskii 
1964, Berezinsky et al. 1990, Gaisser 1991, Stanev 2004, and recent 
reviews by Gaisser \& Stanev 2006, Biermann et al. 2003, 2006, as well 
as Nagano \& Watson 2000) we may be able to learn some of the physics at 
such energies, so perhaps go beyond the LHC.

In this short review we will discuss the latest trends in the quest to 
understand where the extremely high energy particles come from, and how 
we might be able to test our ideas. For lack of space we only give a 
small fraction of all references.

\begin{figure}[h!]
\centering
\includegraphics[bb=0cm 0cm 29.7cm 21cm,viewport=5cm 4.5cm 29.7cm
18cm,clip,scale=0.37]{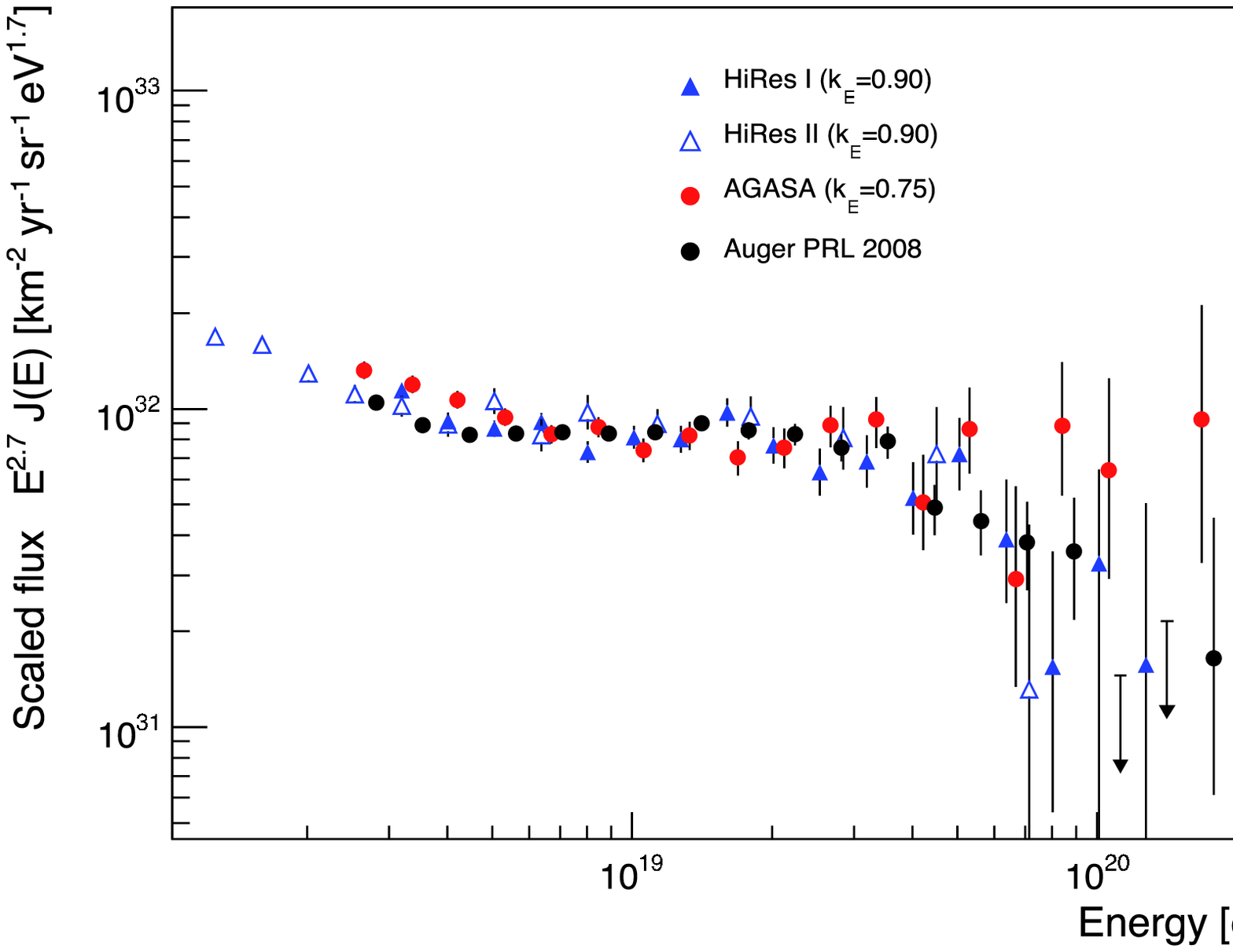}
\caption{This overlay spectrum shows the public data, as of summer 2008, 
of AGASA, HiRes and Auger (Abbasi et al. (HiRes) 2008; Auger 2008b). The 
AGASA event energy has been scaled down by 25 \%, and the HiRes event 
energy has been scaled down by 10 \%. The spectra are adjusted to show a 
common flux near $10^{19}$ eV. The main problem here is the energy 
estimate of the different detectors.}
\end{figure}
\label{UHECR data 2008}

\section{Source candidates}

While very many ideas exist based on detailed physical models for 
possible sources of ultra high energy cosmic ray particles, the best bet 
candidates to explain them are radio galaxies (Ginzburg \& Syrovatskii, 
Blandford, Biermann, et multi al.) and gamma ray bursts 
(M{\'e}sz{\'a}ros, Piran, Rees, Vietri, Waxman, et multi al., with a 
recent summary, e.g., Waxman 2006). As gamma ray bursts are special 
cases of very massive star explosions, their occurrence should correlate 
with galaxies, which have a current starburst, so are strong in the far 
infrared, such as, e.g., M82 (e.g. Kronberg et al. 1985), NGC253, 
NGC2146 and the like; the early models (Biermann 1976, Biermann \& 
Fricke 1977) already allowed the prediction of far-infrared from radio 
fluxes from starburst galaxies, such as NGC2146 (Kronberg \& Biermann 
1981), and so gave a prediction of the supernova rate (today perhaps 
equivalent to a prediction of the gamma ray burst rate). At present the 
statistics of the arrival directions do not support a correlation with 
starburst galaxies. On the other hand, the arrival directions do seem 
compatible with the nearest radio galaxy, Cen A = NGC5128, a source long 
suspected to emit cosmic rays (Ginzburg \& Syrovatski 1963). However, 
this radio galaxy has so little power, it presents a special challenge 
to understand how it could accelerate particles to $10^{20}$ eV, and beyond.

For another nearby radio galaxy, M87 in the Virgo cluster (Ginzburg \& 
Syrovatski 1963, Cunningham et al. 1980), the synchrotron spectrum of 
the knots in the jet has been used to argue that it requires protons at 
$10^{21}$ eV to initiate the cascade in the plasma for scattering the 
non-thermal electrons in order to yield a parameter-free cutoff 
frequency of near $3 \cdot 10^{14}$ Hz, as observed in many knots, hot 
spots and nuclei (e.g. Rieke et al. 1976) of radio emitting active 
galactic nuclei (Biermann \& Strittmatter 1987). This is in fact the 
only argument based on observations which implies the existence of these 
ultra high energy particles in the source. However, it has to be noted, 
that this does not {nec\-es\-sar\-i\-ly} imply that the particles we 
observe come from such sources; it is just plausible for lack of many 
alternatives.

\subsection{Complete samples}

In order to test the idea that radio galaxies are source candidates, we 
have developed the jet-disk symbiosis concept (papers by Falcke et al. 
1995a, b, Markoff et al. 2001, Yuan et al. 2002, Massi \& Kaufman 
Bernardo 2008, etc.). Therefore we need a complete sample of steep 
spectrum radio sources (e.g. teams led by Witzel, see K{\"u}hr et al. 
1981). Table 1 presents such a complete list, differentiated in two sets 
in redshift range; the complete sample takes all extragalactic steep 
spectrum sources down to a flux density limit, which are not already 
known as predominantly starburst galaxies from their far-infrared/radio 
flux density ratio, and that are within the redshift specified. Table 2 
extends the list to slightly higher redshift.

\begin{table*}[htb]
\small
\caption{Properties of the { complete sample} selected in passband 6cm 
(5 GHz), redshift $z\leq 0.018$ and $z\leq 0.0125$ flux density brighter 
than 0.5 Jy, steep spectrum and no starburst, sample of 21 and 14 
candidate sources (Caramete et al. 2008). The distances are corrected 
for the local cosmological velocity field. The FIR/radio ratio can 
readily distinguish radio galaxies from normal galaxies and pure 
starbursts (Kronberg et al., Chini et al.).}
\label{table:1}
\newcommand{\m}{\hphantom{$-$}}
\newcommand{\cc}[1]{\multicolumn{1}{c}{#1}}
\begin{tabular}{|c|c|c|c|c|c|c|c|c|}
\hline
Name&Morphological&Redsh.&Dist.& M$_{BH}$&Core flux
density&B-V&FIR/Radio\\
&type&&Mpc&$10^{8}$M$_{\odot}$&mJy&mag&ratio\\
\hline
\hline
NGC 5128&S0 pec Sy2&0.001825&3.4&2&133361&0.88&3.39\\
NGC 4651&SA(rs)c LINER&0.002685&18.3&0.4&700&0.51&8\\
MESSIER 084&E1;LERG;LINER Sy2&0.003536&16&10&2094.18&0.94&0.17\\
MESSIER 087&E+0-1 pec;NLRG Sy&0.00436&16&31&9480.75&0.93&0.01\\
NGC 1399&cD;E1 pec&0.004753&15.9&3&342&0.95&0.04\\
NGC 1316&(R')SAB(s)00 LINER&0.005871&22.6&9.2&5651.61&&0.06\\
NGC 2663&E&0.007012&32.5&6.1&628.56&&0.08\\
NGC 4261&E2-3;LINER Sy3&0.007465&16.5&5.2&2662.69&0.97&0.02\\
NGC 4696&BCG;E+1 pec LINER&0.009867&44.4&3&518.28&&0.08\\
NGC 3801&S0/a&0.011064&50&2.2&300.25&0.9&0.3\\
IC 5063&SA(s)0+: Sy2&0.011348&44.9&2&321.14&0.93&11.08\\
NGC 5090&E2&0.011411&50.4&7.4&488.13& .. &0.1\\
NGC 5793&Sb: sp Sy2&0.011645&50.8&1.4&51.5&0.79&12.76\\
IC 4296&BCG;E;Radio Galaxy&0.012465&54.9&10&442.22&0.95&0.08\\
\hline
\hline
NGC 0193&SAB(s)0-:&0.014657&55.5&2&285.93&0.98&0.76\\
VV 201&Double galaxy&0.015&66.2&1&450.09&&0.05\\
UGC 11294/4&E0?;HSB&0.016144&63.6&2.9&254.52&&0.33\\
NGC 1167&SA0-;LINER Sy2&0.016495&65.2&4.6&393.09&&0.13\\
CGCG 114-025&SA0-&0.016885&67.4&1.9&443.39&&0.01\\
NGC 0383&BCG;SA0-: LERG&0.017005&65.8&5.5&414.25&&0.21\\
ARP 308&Double galaxy WLRG&0.018&69.7&1&88.54&&0.09\\
\hline
\end{tabular}\\[2pt]
\end{table*}

\subsection{Particle energy and particle flux predictions}

The main indefinite parameter in the jet-disk symbiosis picture is the 
anchoring of the magnetic field at the base of the jet. Spin-down 
powered jets emanating from very near super-massive black holes 
(Blandford \& Znajek 1977, Blandford \& Koenigl 1979, Boldt \& Ghosh 
1999) are one possibility to do this: The jet power is roughly 
proportional to the total radio luminosity (En{\ss}lin et al. 1997), 
especially if we include low power sources in the crude fit. If we 
identify the jet power as an upper limit to the Poynting flux, and use 
the relationship between Poynting flux and maximal particle energy 
(Lovelace 1976, see below), we obtain an expression for the maximal 
particle energy. Furthermore we can assume that the cosmic ray flux is a 
fraction of the total jet power, and so obtain a simple proportionality. 
Calling the mass of the black hole $M_{BH}$, the observed compact radio 
flux density $S_{rad}$ or extended total flux density at 2.7 GHz 
$S_{2.7,tot}$, the luminosity distance $D_L$ to the radio galaxy, the 
maximal particle energy $E_{max}$, and the maximal cosmic ray flux 
$F_{CR}$, we then have here and below:

\begin{equation}
E_{max} \, \sim \, D_L \; S_{2.7, tot}^{1/2}
\end{equation}

and

\begin{equation}
F_{CR} \, \sim \, S_{2.7, tot}
\end{equation}

This flux corresponds to distance attenuation only with $D_{L}^{-2}$.
Interestingly, the mass of the black hole does not even enter here due 
to the simplicity of the Poynting flux argument. Following the argument 
below we might have to multiply the maximum particle energy by 6 - 8 or 
so to simulate the seeding with heavier nuclei from a weak starburst, 
indicated by a relatively large FIR/radio ratio as given in Table 1; 
this ratio is still far below that of a pure starburst, for which it is 
of order 300. However, the numbers given for the maximal energy do not 
include this extra factor. In Table 2 this is indicated by an asterisk.

Accretion powered jets are the other alternative, which works well for 
relatively high current accretion rates (Falcke et al. 1995a, b, 
Ta\c{s}c\u{a}u 2004, Ta\c{s}c\u{a}u et al. 2008):

\begin{equation}
{E_{max}^{\dagger}} \, \sim \, S_{rad}^{1/3} \, D_L^{2/3} \, M_{BH}
\end{equation}

and

\begin{equation}
{F_{CR}}^{\dagger} \, \sim \, S_{rad}^{2/3} \, D_L^{-2/3}
\end{equation}

For distances $<$ 50 Mpc usually NGC5128 = Cen A, possibly NGC1316 = For 
A, and a group around M87 = Vir A dominate in predicted UHECR flux 
(Ginzburg \& Syrovatskii 1963). The first five in flux density of the 
extended flux are ESO137-G006, NGC1316, NGC4261, NGC4486=M87, and NGC5128.

An early attempt to fit older data is shown in Fig. 2.

These two approaches allow to understand the huge range in radio to 
optical flux ratios from active galactic nuclei (Strittmatter \& Witzel 
et al., 1980), and the ubiquity of low flux densities of compact radio 
emission from basically all early Hubble type galaxies (e.g. 
Perez-Fournon \& Biermann 1984). As soon as the accretion rate drops 
below some critical level, spin-down takes over from accretion as the 
powering mode. As pointed out by Blandford the decay time of spin-down 
powered activity is very long, and may allow to understand the 
appearance of ``inverse evolution" for flat spectrum radio sources: It 
just may be the growing number of ``old" central activity since the 
activity per comoving starburst and central activity in galaxies peaked 
in the redshift range 1.5 to 2. The activity decreased by about a factor 
of 30 since then, and so we have an increasing population of early 
Hubble type galaxies, which had their prime activity years some long 
time ago. If all central black holes stay active - as observations 
suggest, albeit at a very low level - then a subset of the population of 
these black holes will aim their jet at Earth, and so give rise to a 
weak, but dominant flat radio spectrum. One consequence is that most 
central supermassive black holes should have close to maximal spin.

All these weakly active galactic nuclei will also accelerate particles 
to high energy, but have a flux, which is generally extremely low. Such 
particles would have to be injected from the interstellar medium of the 
early Hubble-type host galaxy, and can be argued to be mostly protons, 
with some Helium. Their maximal energy will be relatively low.

\begin{figure}[h!]
\centering
\includegraphics[bb=0cm 0cm 21cm 30cm,viewport=1cm 1.0cm 21cm
14.7cm,clip,scale=0.4]{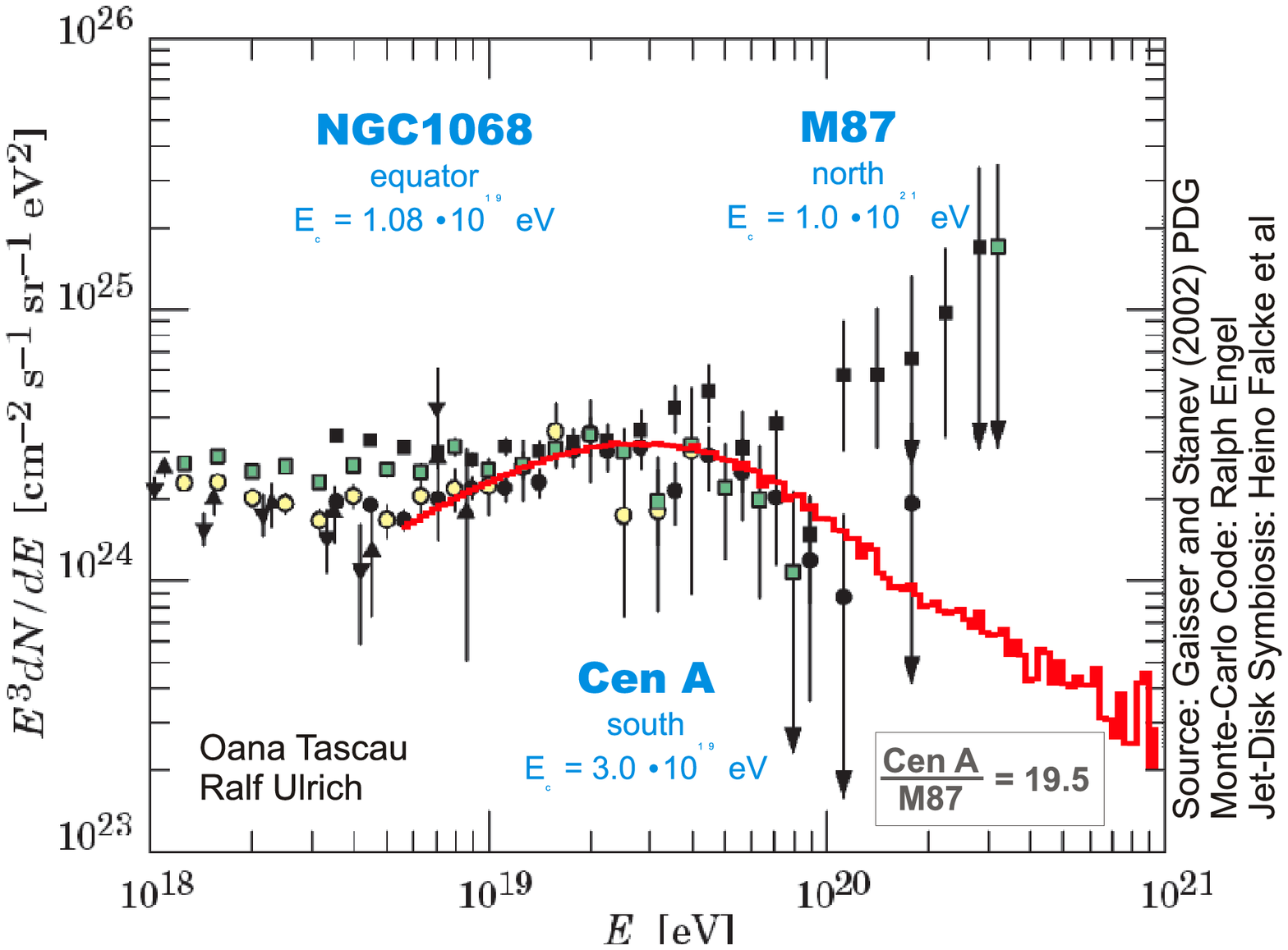}
\caption{This spectrum shows a best fit, including only three sources, 
NGC1068, Cen A, and M87. The fit was achieved by setting the ratio of 
the flux of Cen A relative to M87 to 19.5 at $10^{19}$ eV; key to the 
match was the lower maximal energy of Cen A - at highest energy M87 is 
still the strongest: The cutoff is due to source limits, not due to 
GZK-interactions. The flux of NGC1068 is at 0.7 relative to M87, and its 
maximal energy is only $10^{19}$ eV; relative to Cen A, this is 
insignificant here. This is quoted from the M.Sc. thesis of O. 
Ta\c{s}c\u{a}u (2004).}
\end{figure}
\label{Oana Spectrum}

Table 2 gives the predictions.

\begin{table*}[htb]
\small
\caption{{ UHECR predictions:} Using core flux-density at 5 GHz for the 
complete sample of 29 steep spectrum sources (K{\"u}hr et al., 1981). 
Col. 4: (*) Core flux density estimated from the total flux density by 
using $\log(P_{core})=11.01+0.47\log(P_{tot})$, cf. Giovannini 1988; 
Col. 5 \& 6: Relative values of the particles maximum energy and UHECR 
flux by using spin-down (equations above). Col. 7 \& 8: ($\dagger$) 
Relative values of the particles maximum energy and UHECR flux by using 
accretion (O. Ta\c{s}c\u{a}u). These predictions do not take into 
account losses, these numbers just reflect the spatial limit, and the 
flux reduction with distance squared. Energies with an asterisk may have 
to be increased due to weak starburst seeding of heavier elements; this 
could be an order of magnitude)}
\label{table:1}
\newcommand{\m}{\hphantom{$-$}}
\newcommand{\cc}[1]{\multicolumn{1}{c}{#1}}
\begin{tabular}{| c |c | c | c | c |c | c | c |}
\hline
Source & D & M$_{BH}$ & $S_{5GHz}$ & $ E_{max}/E_{max}^{M87} $ &$ 
F_{CR}/F_{CR}^{M87} $& $ {E_{max}/E_{max}^{M87}}^{\dagger} $ &$ 
{F_{CR}/F_{CR}^{M87}}^{\dagger} $\\ & (Mpc)& ($\times 10^9 M_\odot 
$)&(mJy)& & & & \\
\begin{small}(1)\end{small}&\begin{small}(2)\end{small}&\begin{small}(3)\end{small}&\begin{small}(4)\end{small}&\begin{small}(5)\end{small}&\begin{small}(6)\end{small}&\begin{small}(7)\end{small}&\begin{small}(8)\end{small}\\[0.5ex]
\hline
ARP 308&69.7&0.1&88.53*& 0.72&0.027&0.03&0.04\\
CGCG 114-025&67.4&0.19&2260& 0.80&0.036&0.15&0.33\\
ESO 137-G006&76.2&0.92&631.32*& 1.79&0.12&0.51&0.13\\
IC 4296&54.9&1&214& 0.49&0.026&0.31&0.08\\
IC 5063&44.9&0.2&321.15*& 0.23 *&0.0067&0.06&0.12\\
NGC 0193&55.5&0.2&285.93*& 0.34&0.010&0.07&0.09\\
NGC 0383&65.8&0.55&414.25*& 0.70&0.029&0.24&0.11\\
NGC 1128&92.2&0.2&280.2*& 1.1&0.036&0.1&0.07\\
NGC 1167&65.2&0.46&393.1*& 0.42&0.011&0.2&0.1\\
NGC 1316&22.6&0.92&26& 1.3&0.82&0.08&0.03\\
NGC 1399&15.9&0.3&10& 0.11&0.012&0.01&0.02\\
NGC 2663&32.5&0.61&160& 0.22&0.012&0.12&0.09\\
NGC 3801&50&0.22&635& 0.25&0.0063&0.09&0.17\\
NGC 3862&93.7&0.44&1674& 0.97&0.027&0.39&0.21\\
NGC 4261&16.5&0.52&390& 0.34&0.11&0.09&0.26\\
NGC 4374&16&1&168.7& 0.18&0.033&0.13&0.15\\
NGC 4486&16&3.1&2875.1&1&1&1&1\\
NGC 4651&18.3&0.04&15& 0.12 *&0.012&0&0.03\\
NGC 4696&44.4&0.3&55& 0.37&0.018&0.05&0.04\\
NGC 5090&50.4&0.74&268& 0.50&0.026&0.23&0.1\\
NGC 5128&3.4&0.2&6984& 0.43*&4.0&0.04&3.63\\
NGC 5532&104.8&1.08&194.58*& 0.98&0.023&0.5&0.05\\
NGC 5793&50.8&0.14&95.38*& 0.27 *&0.0072&0.03&0.05\\
NGC 7075&72.7&0.25&20& 0.34&0.0054&0.04&0.01\\
UGC 01841&84.4&0.1&365.46*& 1.2&0.053&0.05&0.08\\
UGC 02783&82.6&0.42&541& 0.40&0.0058&0.23&0.11\\
UGC 11294/4&63.6&0.29&314& 0.35&0.0075&0.11&0.09\\
VV 201&66.2&0.1&450.1*& 0.82&0.040&0.04&0.11\\
WEIN 045&84.6&0.27&321.6*& 0.98&0.034&0.13&0.08\\
\hline
\end{tabular}\\[2pt]
\end{table*}

\subsection{Scattering model}

Basic questions on the effect of intergalactic and galactic magnetic 
fields on the propagation of ultra high energy charged particles are 
whether a) is there (almost) no effect, b) is a systematic bending of 
orbits key or c) is there a general scattering (see, e.g., Das et al. 
2008). Any systematic shift is not apparent at the present time with the 
sparse data, while a general scattering seems required. There may be a 
general systemic shift, which would provide a location-dependent change 
of direction for the least scattered events from anyone source. For lack 
of strong evidence we ignore such a plausible shift for the moment.

Since the data suggested a near isotropic sky distribution in 1995 
(Stanev et al. 1995), and a more correlated distribution with more and 
homogeneous data (Auger-Coll. 2007, 2008a) a scattering model is 
suggested which spreads arriving events almost evenly; an alternative 
would have been to have many sources, but no such model is currently 
plausible. A simple single scattering plasma physics approximation 
suggests a scattering model of $\theta^{-2}$ in scattering angle 
$\theta$ per solid angle, which spreads events evenly into logarithmic 
rings $\Delta \theta / \theta \, = \, const$ (Curu\c{t}iu et al. 2008). 
The detailed magneto-hydro-dynamic (MHD) simulations of Das et al. 
(2008) support a description as a simple such power-law at high energy, 
while at low energy the spreading is smoother and broader, corresponding 
to multiple scattering. For simplicity we use here $\theta^{-2}$ with a 
core of 3 degrees, and a maximum of 90 degrees; the core is to reflect 
what happens in the galactic disk (Beuermann et al. 1985, Snowden et al. 
1997), while the general scattering distribution may reflect either 
scattering in the cosmological magnetic fields as in Das et al. (2008), 
or scattering in a galactic magnetic wind halo (Parker 1958, 
Simard-Normandin \& Kronberg 1980, Parker 1992, Ahn et al. 1999, Hanasz 
et al. 2004, Westmeier et al. 2005, Chy\.zy et al. 2006, Breitschwerdt 
2008, Kulsrud \& Zweibel 2008, Caramete et al. 2008). The main 
difference in these two sites of scattering is that only for scattering 
by cosmological magnetic fields we obtain appreciable delay times, 
changing the spectrum (Stanev et al. 2003, Das et al. 2008). We neglect 
here the possibility that the source itself might be appreciably 
extended, as Cen A is, with a 10 degree size already in sensitivity 
limited data (Junkes et al. 1993). We also do not take into account the 
effect of the local shear flow, dragging magnetic fields along (Kulsrud 
et al. 1997, Ryu et al. 1998, En{\ss}lin et al. 1998, Kronberg et al. 
1999, Gopal-Krishna et al. 2001, Ryu et al. 2008), in the cosmological 
filament around Cen A; the shear flow is expected to be parallel to the 
outer shape of the radio source. This shear flow can be expected to 
scatter particles, making the sites of origin appear correlated with the 
large scale filament.

With such a simple prescription we can turn a source list with predicted 
cosmic ray fluxes into probable sky distributions (Caramete et al. 
2008). We used sets of 100 simulated events each, and performed $10^{6}$ 
such Monte-Carlo runs, for a total of $10^{8}$ simulated events: Using 
the predicted fluxes, and the scattering distribution we sample the 
entire list of Table 2 out to a given redshift (we used 0.0125, 0.018, 
and 0.025) not taking here into account the GZK-attenuation. Including 
the sky sensitivity for both Auger and HiRes we again generate detected 
simulated sets of events of 100 each, and we then find with the 
predictions listed above, that using the V{\'e}ron-catalogue 
(V{\'e}ron-Cetty \& V{\'e}ron 2006) as procedure (Auger-Coll. 2007, 
2008a) in searching for correlations we get a broad probability 
distribution around 50 percent of correlated events in the Auger sky, 
and about 30 percent in the HiRes sky (Curu\c{t}iu \& Caramete 2008). We 
also find a relatively large ratio between the number of simulated 
events in the Auger-sky versus the HiRes-sky. We find a larger predicted 
number of events in the HiRes-sky versus the Auger-sky only for galaxies 
selected to represent a parent population of gamma ray bursts (i.e. 
selected at 60 $\mu$). All this just reflects the well-known fact, that 
the sky is not homogeneous in the nearby universe.

\begin{figure}[h!]
\centering
\includegraphics[bb=0cm 0cm 21cm 30cm,viewport=1.3cm 1.3cm 20cm 
13cm,clip,scale=0.42]{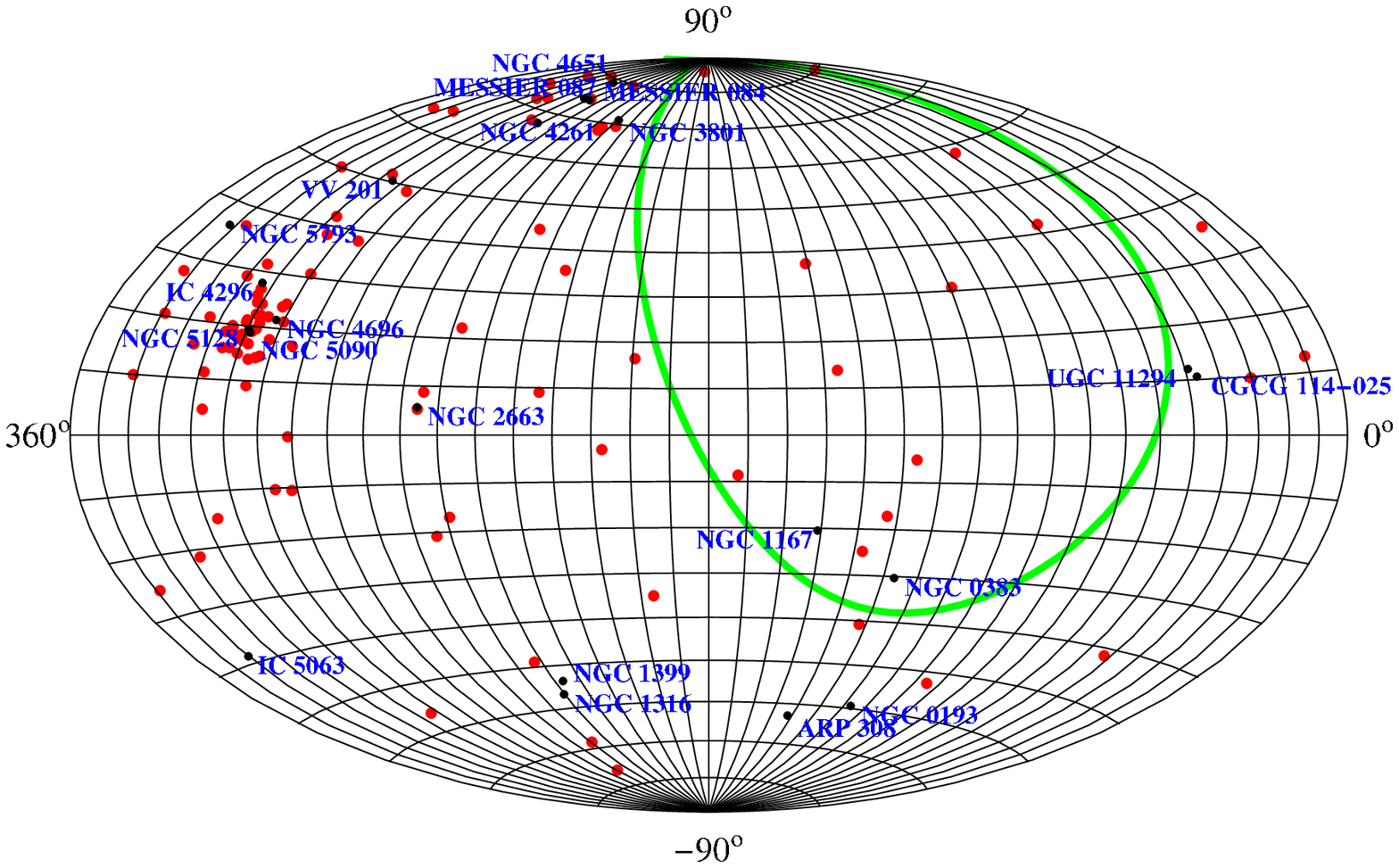}
\caption{Aitoff projection in galactic coordinates of the selection from 
the NASA/IPAC Extragalactic Database (NED) in passband 6cm (5 GHz), 
redshift $z\leq 0.0125$ flux density brighter than 0.5 Jy, steep 
spectrum and no starburst, sample of 14 candidate sources and 100 
virtual events from these sources (Curu\c{t}iu \& Caramete 2008) using 
the core of 3 degrees distribution of scattering and weighted 
contribution from the accretion model (O. Ta\c{s}c\u{a}u). The green 
line highlights the area of the sky not visible from the Auger site.}
\end{figure}
\label{SkyPlot}

We note that gamma ray bursts have been predicted to show only protons 
(Rachen \& M{\'e}sz{\'a}ros 1998), as end-products from decaying 
neutrons, the only particles that may escape from magnetic confinement 
before adiabatic losses set in; the neutrons are believed to be created 
in proton-$\gamma$ collisions, so arise from regions of very high photon 
density. The HiRes data on air-fluorescence are consistent with such a 
picture (Talk by Sokolsky 2008, ISVHECRI meeting).

However, this does not easily explain the cloud of events in the Auger 
data around the obvious radio galaxy Cen A, of which 5 at least are 
directly confined within the outlines of the radio emission (Junkes et 
al. 1993, Rachen 2008).

\subsection{Determining anisotropy}

One major question with the sparsity of data is how to determine a 
measure of anisotropy quantitatively. The astronomical sky shows two 
extreme measures directly: The microwave background, once corrected for 
the dipole anisotropy to to our peculiar velocity is as perfect as one 
could imagine (Komatsu et al. 2008). On the other hand, the nearby 
distribution of galaxies, out to at least 300 Mpc shows anisotropy. 
Different classes of galaxies have different measures of anisotropy, and 
the most anisotropic are the galaxies which harbor very large 
super-massive black holes, giant elliptical galaxies.

Therefore, clearly the best measure of anisotropy is to determine for a 
given set of ultra high energy cosmic ray arrival directions, where in 
the range from perfect isotropy to maximal anisotropy, the set of nearby 
very large supermassive black hole host galaxies, these events lie. 
Clearly, as demonstrated already, the arrival directions are somewhere 
in between - assuming of course, that they relate to astrophysical and 
known object classes.

\section{Problems with radio galaxies}

\subsection{The Poynting flux limit}

As Lovelace (1976) has originally shown, the Poynting flux is a lower 
limit to the energy flux along a relativistic jet, and can be written as 
basically proportional to the maximal particle energy containable 
squared. The numbers are such, that for particles reaching $10^{21}$ eV, 
$10^{47}$ erg/s is a conservative lower limit:

\begin{equation}
L_P \; = \; \frac{B^{2}}{4 \pi} \pi \theta^{2} z^{2} c
\end{equation}

\noindent and

\begin{equation}
E_{max} \; = \; Z e B \theta z
\end{equation}

\noindent which implies

\begin{equation}
L_P \; = \; 10^{47} \, {\rm erg/s} \, \left(\frac{E_{max}}{Z \, 10^{21} 
\, {\rm eV} }\right)^{2}
\end{equation}

M87 and Cen A have energy flows along the jet of order $< \, 10^{45}$ 
erg/s, $< \, 10^{43}$ erg/s (Whysong \& Antonucci 2003), respectively. 
This implies that it is completely impossible for the Cen A jet to 
supply the environment to accelerate protons to $> \, 10^{20}$ eV, but 
allowing $Z \, > 1$ changes this conclusion. A shock in upstream flow 
with shock Lorentz facor $\gamma_{sh}$ (Gallant \& Achterberg 1999) adds 
another factor, and finally intermittency $f_{flare} < 1$ also helps, so 
visible directly in Her A (Gizani \& Leahy 2003, and Nulsen et al. 
2005). We finally obtain

\begin{equation}
L_P \; = \; \frac{c}{4 \pi} \, f_{flare} \, \left(\frac{E_{max}}{e \, Z 
\, \gamma_{sh}}\right)^{2}
\end{equation}

The discrepancy is so large, that perhaps all three elements, heavy 
elements, relativistic shocks, and intermittency or flaring, are 
required; only pure Fe at the highest particle energies has a chance for 
Cen A to do it all by itself. In Cen A there is clearly a starburst 
happening, a phase of strongly enhanced star formation and supernova 
activity, in which the local cosmic rays can be expected to be 
substantially increased. The heavier elements as seeds of ultra high 
energy cosmic rays are therefore perhaps plausible, since it is much 
faster to accelerate particles from the knee of cosmic rays, where 
Carbon, Oxygen, Neon to Sulfur are important (Stanev, Biermann, \& 
Gaisser 1993), as was shown by Gallant \& Achterberg in a different 
context (1999).

Let us consider the approach of Gallant \& Achterberg (1999) in more 
detail so understand, what it would lead to: A young starburst has 
injected a strong population of galactic cosmic rays, still in the 
spectral injection limit, and now a very powerful highly relativistic 
shock driven by a jet plows right through this environment. The 
starburst was visibly triggered by a merger between two galaxies, 
probably both with super-massive central black holes, and when the two 
black holes finally also merge, orbital spin wins and induces a 
spin-flip of the final black hole relative to the spin direction of the 
previously more massive single black hole (e.g. Gergely \& Biermann 
2007, 2008, also see below). Therefore the newly powered jet plows 
through material untouched by the previous older jet, and just filled 
with interstellar medium, highly excited by the all the explosions of 
very massive stars, most importantly Wolf-Rayet stars. Wolf-Rayet stars 
render all the heavy element and Helium cosmic ray particles (Stanev et 
al. 1993). Therefore we are considering the energy gains of a highly 
relativistic particle of energy $E_{1}$, going back and forth across a 
relativistic shock, with shock Lorentz factor $\Gamma_{sh}$, gaining 
energy each cycle time (e.g., Drury 1983). Gallant \& Achterberg show 
that the initial energy jump is by a factor of $\Gamma_{sh}^{2}$, and 
all subsequent energy jumps are just by about a factor of $2$.

\begin{equation}
E_{2} \, = \, E_{1} \, \Gamma_{sh}^{2} \, 2^{n}
\end{equation}

\noindent where $n$ is the number of subsequent cycles. This can be 
rewritten with $\epsilon \, = \Gamma_{sh}^{2} \, 2^{n} \, >> 1$ as

\begin{equation}
E_{2} \, = \, E_{1} \, \epsilon
\end{equation}

\noindent which turns a spectrum of

\begin{equation}
N_{0} \, {\left(\frac{E}{E_{0}}\right)}^{-p} \, d E
\end{equation}

\noindent into

\begin{equation}
\frac{N_{0}}{\epsilon} \, {\left(\frac{E}{E_{0} \, 
\epsilon}\right)}^{-p} \, d E
\end{equation}

\noindent which implies that a spectrum is shifted in flux down by a 
factor of $\epsilon$, and also over in energy by the same factor.

Considering then the seed population of energetic particles at the knee 
(Stanev, Biermann, \& Gaisser 1993) this implies, that the spectral 
differentiation by $Z$ at the knee of cosmic rays is shifted over, and 
down by another factor of $\epsilon$, so reproducing the spectral 
structure in $Z$. So, given the spectral bending of the various elements 
at the knee, we can readily predict the spectral shapes of the spectrum 
in energy per particle, with the elements like Carbon, Oxygen etc first, 
shifting ultimately to Iron.

We can obviously check whether the energy jump required, by about 1000 
to 3000, is sensible. For the shock Lorentz factor we can take some 
value between $10$ and 50 (Begelman et al. 1994, 2008, Gopal-Krishna et 
al. 2004, Miller-Jones et al 2004, Ghisellini \& Tavecchio 2008, et 
multi al.), and we then estimate the number of subsequent cycles required:

\begin{equation}
1000 \; {\rm to} \, 3000\, = \, \Gamma_{sh}^{2} \, 2^{n}
\end{equation}

For a shock Lorentz factor of 10 this requires $n$ from 3 to 5, and for 
a shock Lorentz factor of 50 it requires no extra jump at all. 
Obviously, $n = 0$ would minimize the smearing in energy during the 
shift up in energy.

In this speculative model the knee structure (Stanev et al. 1993) in 
chemical composition and spectrum is preserved at very high energy. 
Super-luminal shocks can squeeze this overall spectral structure 
(Hoffmann \& Teller 1950), and may deplete it at lower energies, but 
will basically still preserve it (Meli 2008, see below). As only a small 
fraction of all Wolf-Rayet stars turn into gamma ray bursts, the cosmic 
ray contribution from gamma ray bursts to the seed population is likely 
to be small (Pugliese et al. 2000). As was noted by Biermann (1993) and 
Stanev et al. (1993), there is an accentuation at the knee, the polar 
cap component with a $E^{-2}$ spectrum, now probably detected in its 
loss limit of cosmic ray electrons by the ATIC experiment (Chang et al. 
2008). This polar cap component sharpens the knee features of each 
element; during the strong jump in energy from the knee up there will be 
some inevitable smearing, but this polar cap component will help keep 
the features visible.

Of course, very much later in the evolution of an activity episode these 
seeds will be replaced by the normal average chemical abundances, normal 
for an elliptical galaxy as typical host for a radio galaxy with perhaps 
inflow from the local intergalactic medium.

\subsection{Magnetic fields in jets}

The radio polarization data (e.g. Bridle \& Perley 1984) strongly 
suggest that the magnetic field decreases with distance squared from the 
central black hole, just as in the Solar wind along the rotational 
symmetry axis (Parker 1958). A magnetic field decaying as distance 
squared along the jet would never allow enough space for ultra high 
energy particles to be accelerated. However, highly oblique shocks could 
mimick such a pattern (Becker \& Biermann 2008) even for a basic 
magnetic field oscillating around a inverse linear decay along the jet. 
In such a case, the magnetic field could be strong enough far along the 
jet to allow the acceleration of ultra high energy particles (Hillas 1984).

\subsection{Spectral limit}

All these suggestions above lead to another discrepancy, considering the 
likely spectrum of energetic particles: the radio data suggest a typical 
spectrum of $E^{-2.2}$ (Bridle \& Perley 1984), and only rarely a 
spectrum as flat as $E^{-2.0}$, while fitting the lower energies of the 
ultra high energy cosmic ray spectrum suggests possibly even $E^{-2.7}$ 
(Berezinsky et al. 2006). On the other hand, the observed flux of ultra 
high energy particles is already so high, that a simple straight 
continuation of the spectrum $E^{-2.2}$ versus $E^{-2.0}$ would imply an 
extra factor of about 200 in required energy flux.

However, the phenomenon of incomplete Comptonization (Katz 1976) leads 
us to ask, whether an analogy of relativistic particles to photons might 
be possible: Photons can show a diminished low energy spectrum in cases, 
when the number of photons is constrained independently of its energy 
content, leading to a finite chemical potential describing what might be 
called a starved spectrum. Such spectra were crucial to understand the 
X-ray spectra of active X-ray binary stars (see also Katz, Lightman, 
Sunyanev, et multi al.). It appears that starved cosmic ray spectra are 
also possible, as first explorations (Meli 2008) show, that the 
combination of a subluminal shock with a superluminal shock (Hoffmann \& 
Teller 1950, Drury 1983), a natural reconfinement shock arrangement 
(Mach 1884-1898, R. Sanders 1983, M. Norman et al., T. Jones et al.), 
would indeed lead to spectra with a dearth of low energy particles. This 
would lower the energy requirement considerably. This may actually be 
required for spectra as steep as $E^{-2.2}$ or steeper.

\subsection{Neutrons}

In analogy to gamma ray bursts Rachen (2008) has suggested also for 
radio galaxies to accelerate protons to high energy, then transforming 
them in p-$\gamma$-collisions to neutrons (Puget et al. 1976, Rachen \& 
M{\'e}sz{\'a}ros 1998) to get them out at high energy without adiabatic 
losses. Given that the jet in Cen A is apparently not close to the line 
of sight, this could be tested for consistency, if the arriving events 
interpreted as original neutrons were linearly arranged sorted by 
particle energy, with about 16 degrees in the plane of the sky at 100 
EeV, or less. The sparse data do not contradict this; however, as noted 
above, this environment may not be conducive to the acceleration of 
protons to extremely high energy. It has to be noted that neutrons at 
300 EeV (the observed maximum energy, Fly's Eye 1993) could travel 
straight from Cen A to Earth, a distance of 3.4 Mpc, with only a small 
fraction decaying back to protons.

\subsection{The three horizons}

Observed protons which come from large distances are diminished in 
energy by interaction with the microwave background (Greisen 1966, 
Zatsepin \& Kuzmin 1966) for energies beyond about $6 \cdot 10^{19}$ eV; 
nuclei suffer from photo-dissociation (Rachen 1996, Stecker, \& Salamon 
1999, Hooper et al. 2007, 2008, Allard et al. 2008). This leads to the 
GZK-horizon, which is strongly dependent on the energy of the particle 
arriving at Earth; if protons at $> \; 6 \cdot 10^{19}$ eV, about half 
of the events should come from less than 50 Mpc, and close to 90 percent 
should come from less than 200 Mpc. Enhancing this line of reasoning, 
there is obviously for each element and isotope separately a horizon, 
from which this specific element has a good chance of surviving 
photo-dissociation. It could be interesting to investigate the paths in 
the charge-mass (Z, A)-diagram, the nuclei take, and how often they just 
disintegrate on their own, sowing the environment with decay products; 
this is a concept just the reverse of the nuclear element build-up 
(Burbidge et al. 1957). Another query is to understand to what degree 
these processes might already happen inside the relativistic radio jet. 
And a third investigation might center on the spallation products among 
the seed population resulting from the ubiquitous nuclear collisions 
happening in the dense environments of Wolf-Rayet stars after they blow 
up, and before they get hit by the relativistic jet; do we have a chance 
to discern these spallation products, like the sub-Fe elements or the 
Li, Be, B nuclei, among the ultra high energy cosmic ray particles?

These particles may not come from arbitrarily large distances due to 
magnetic scattering (Stanev et al. 2003, Das et al. 2008). This is the 
magnetic horizon. In the MHD simulations of Das et al. (2008) this is at 
100 Mpc at $>$ 60 EeV for protons. Due to the chain of 
photo-dissociation and the ensuing modification of the nuclear charge, 
this horizon is strongly dependent on the path, the nuclei take upon 
interaction.

In the search for directional correlations on the sky we require the 
large scale structure scales, and this is $\geq$ 300 Mpc (Peebles 1989, 
Rudnick et al. 2007). So directional correlations are expected (see 
Tinyakov et al., Tkachev et al., \& Finley \& Westerhoff 2004, Mari\c{s} 
2004, Caramete et al. 2008, et multi al.) up to the corresponding 
redshift to be far more common than by chance. This is the correlation 
horizon.

\subsection{The HiRes vs. Auger discrepancy}

The HiRes collaboration (2008b) has disputed the Auger (2007, 2008a) 
result that the arrival directions of ultra high energy cosmic rays are 
correlated with active galactic nuclei in the V{\'e}ron catalogue 
(V{\'e}ron-Cetty \& V{\'e}ron 2006). The authors emphasize that this 
catalogue is incomplete, and so we are using it only in the exact sense 
in which the original Auger publication is using it, as an instrument of 
comparison. HiRes finds less than random correlations. We noted already 
above, that there are fewer such correlations expected in the North from 
a simple simulation of arriving events from radio galaxies (on average 
1/3 vs. 1/2); since we are using here very small number statistics, we 
may not have to look any further. Another effect might play an 
additional role: As magnetic scattering increases rapidly with lower 
particle energy (Das et al. 2008), one might speculate that in the HiRes 
sample the uncertainty of energy determination at the low energy 
threshold might be large enough to add additional smearing of directions 
due to magnetic fields, and so decreasing any coincidental directional 
correlation. Obviously also, the final energy calibration of HiRes 
versus Auger is a remaining serious issue.

As already noted, just using a simple scattering model and the notion 
that radio galaxies are the sources predicts that about half the events 
should be correlated in the procedural sense for the Auger sky, in the 
limit of large numbers.

\subsection{Application to Cen A}

Taking all these ideas together suggests that maybe we require all of 
the four concepts mentioned above at the same time, relatively heavy 
elements (perhaps Carbon and Oxygen at somewhat lower energies, and Iron 
at higher energies), flaring, starved particle spectra, and weakly 
relativistic shocks. And in addition there may be a subset of pure 
protons from neutron decay at the lower energies.

However, using Cen A as the main source engenders another problem: The 
MHD simulations of Das et al. (2008) suggest strongly a scattering 
distribution of a power-law at high energy, for protons only. If we 
argue that heavier {ele\-ments} are the key, then these magnetic fields 
are either much weaker, or much more structured (more structure weakens 
the scattering, for a given total energy per large volume). In the 
magnetic field data in our galaxy (Beck et al. 2003) there is already 
strong evidence for small scale substructure, since different measures 
of the magnetic field yield very different numbers: {lin\-e\-ar} 
measures such as Faraday Rotation Measures indicate much lower strengths 
of the magnetic field than quadratic measures such as synchrotron 
emission. This is typical for small scale substructure (H. Lee at el. 
2003, Avillez \& Breitschwerdt 2004), where for a given total energy 
content high intensity sheets can hold all the energy for a small volume 
fraction; in such a picture linear measures give a much smaller number 
than quadratic measures, as is well known from mathematically isomorphic 
arguments in thermal emission. Of course we should be comparing the 
proper integrals, also involving the spatial distribution of thermal 
electron density and cosmic ray electron density; we ignore all this in 
our simple didactic exercise.

We can quantify this by integrating along a long thin cylinder of unit 
length; we refer to the magnetic field as $B_{0}$, when it is 
homogeneous, and for the inhomogenous case the magnetic field is $B_{1}$ 
over most of the length, and enhanced by a factor $1/x$ in a region of 
length $x$: This then gives for the integrated energy density

\begin{equation}
B_{1}^{2} \times \frac{1}{x} + B_{1}^{2} \times (1 - x) \; = \; B_{0}^{2}
\end{equation}

\noindent where we keep the integrated energy content $B_{0}^{2}$ 
constant. The linear measure of the magnetic field is then given by

\begin{equation}
B_{1} \times \frac{1}{x} \times x + B_{1} \times (1 - x) \; = B_{1}
\times (2 - x)
\end{equation}

We now vary $x$ to see how the linear measure varies with $x$, keeping 
the entire energy content fixed.

Combining the first equation with the second yields

\begin{equation}
\sqrt{\frac{x}{1 - x}} \times (2 - x)
\end{equation}

for the ratio of linear measure versus quadratic measure. In the limit 
of small $x$ this is simply $\sqrt{x}$. The observations suggest that 
this ratio is of order 1/5, and so $x = 0.04$ by order of magnitude. 
This implies that most of the magnetic energy is contained in shells of 
a volume a few percent, possibly as low as 1 percent. Since the linear 
measure is proportional to the bending of ultra high energy cosmic rays, 
this implies that the bending is reduced by a factor between 5 and 10 
over what we might reasonably expect otherwise.

Using the approach of Cox (1972) with the environment of the tenuous hot 
phase of the interstellar medium (Snowden et al. 1997) the cooling stage 
of an expanding shell of a supernova remnant might lead to such a 
configuration, of a very thin shell at large distances, with strong 
magnetic fields. In such a picture this stage would encompass most of 
the supernova's energy dissipation, and so similar considerations may 
apply to the interpretation of the X-ray data (Snowden et al. 1997).

Begelman (1995) has shown that an analogy of supernova remnants to radio 
galaxies can be illuminating, and so might also lead to thin shells of 
high magnetic fields, in turn decreasing the scattering of ultra high 
energy particles in intergalactic space.

If both of these applications were realized in Nature, magnetic 
scattering of nuclei of $Z >>1$ might appear similar to scattering of 
protons in environments without allowing for such fine sub-structure 
(Stanev 1997, Stanev et al. 2003, Armengaud et al. 2005, Dolag et al. 
2005a, b, Ryu et al. 1998, 2008, Takami \& Sato 2008, Das et al. 2008); 
taking the interstellar medium data literally and also the power-law 
scattering found by Das et al. (2008), would suggest very tentatively, 
that $Z \simge 6$ is quite plausible. This implies that protons would be 
scattered very little. And so, the fact that very few events directly 
point to plausible sources, except at Cen A (protons are energetically 
not plausible from this source, see above), implies under these 
assumptions, that the proton fraction among the events must be small.

However, such considerations on the small scale structure of magnetic 
fields both in the interstellar as well as intergalactic medium must 
remain speculation at this time.

\subsection{Merging black holes}

There is evidence that each episode of an active galactic nucleus is 
triggered by the merger of the host galaxy with another galaxy. In a 
major merger the second galaxy will also have a central super-massive 
black hole, and so the
merger of the two black holes will follow, leading to a spin-flip: The 
spin axis of the final merged black hole will differ from the spin axis 
of the preceding more massive black hole. For the characteristic mass 
ratio range $3\div 30$ of the merging super-massive black holes 
(inferred from the Press-Schechter mass distribution of galaxies, Press 
\& Schechter 1974, as well as observations) the occurrence of the 
spin-flip was shown to be caused by the superposition of the spin-orbit 
precession and energy dissipation due to gravitational radiation 
(Gergely \& Biermann 2007, 2008). This effect can be observed through 
first a sweeping of the jet (Gopal-Krishna et al. 2003), and then a 
switch in jet direction (Rottmann 2001; Zier \& Biermann 2001, 2002; 
Merritt \& Ekers 2002).
In this context it is an unsolved question, how it is possible that 
super-massive black holes undergo many mergers and keep their spin high 
at the same time, as typically a merger reduces the spin (Hughes \& 
Blandford 2003, Berti \& Volonteri 2008). However, differential 
dynamical friction during the spiraling down of an incoming black hole 
and its accompanying core of its host galaxy may lead to a partial 
alignment, if the receiving more massive core is co-rotating with its 
black hole (Gergely \& Biermann).

\subsection{High energy neutrinos}

Following a spin-flip the jet has to carve out a new channel in the 
surrounding material, strongly enhanced possibly due to the preceding 
merger. This will result in powerful shock waves, as the jet plows 
through this environment. This in turn will lead to extreme particle 
acceleration, and strong interaction, as the molecular clouds become the 
near-perfect beam-dump. Furthermore, the first strong shock in the jet 
can accelerate particle in an environment with fairly high photon 
density, either from the accretion disk, or from the emission of the jet 
itself, and so produce lots of high energy neutrinos. The last strong 
shock in the jet, where it goes subsonic, or sub-Alfv{\'e}nic, or stops 
altogether, will produce the high energy particles, that can most 
readily escape, and so perhaps make up those particles which we observe 
(Becker \& Biermann 2008). The very rare but most powerful sources, the 
Fanaroff-Riley class II radio galaxies will be sources of ultra high 
energy cosmic rays and high energy neutrinos, but none seems close 
enough to our Galaxy to be a detectable source at high energy for 
particles. In summary we predict that most neutrinos will be detected 
from flat spectrum radio sources such as BL Lac type sources (just those 
Fanaroff-Riley class I sources aiming at us with their relativistic 
jets), while the observed ultra high energy cosmic rays may come 
predominantly from Fanaroff-Riley class I radio galaxies as well as BL 
Lac type AGN.

However, there is a difficulty, should it be true, that some, perhaps 
many, of the ultra high energy cosmic ray particles are nuclei such as 
Carbon, Oxygen, or even heavier, as conclusively argued above for the 
case of Cen A (see, e.g., Anchordoqui et al. 2008). In that case, any 
interaction of nuclei with a photon field yields just a 
photo-dissociation, and a reduced flux of neutrinos. We obtain neutrinos 
only in a second interaction, with the nucleon split off in this first 
interaction again interacting with the photon field. So, in this case, 
the combined probability for such an interaction is the product of the 
optical depth for photo-dissociation $\tau_{1}$ with the further optical 
depth for p-$\gamma$-interaction $\tau_{2}$. Now, we have also argued 
and demonstrated with an example, Her A, above, that active galactic 
nuclei do most of their interesting activity in a flaring mode. In a 
flaring mode, it is readily expected (applying the equations and numbers 
in Becker \& Biermann 2008), that both optical depths attain values 
above unity, and so from nearly no neutrinos we predict in strong flares 
a huge flux of neutrinos. If this expectation is borne out, the 
detection would also be much easier against the atmospheric neutrino 
background.

\section{Future}

Since the chemical composition enters here at four points of reasoning, 
we probably require a ``principal component analysis", fitting at once 
1) the air fluorescence data, 2) the scattering distribution (note that 
scattering angles of more than 90 degrees are plausible even for sources 
as near as Cen A: Das et al. 2008), including a possible systematic 
shift of the core of the distribution, 3) the delay time distribution, 
which enters the microwave background interaction for protons and in 
photo dissociation for nuclei, and 4) the magnetic horizon, out to which 
we can receive ultra high energy particles from sources.

The task is to predict a chemical composition and associated spectrum 
for a source, then propagate all nuclei and the protons through another 
prediction of the cosmic web of magnetic fields with all its un-known 
fine-structure, include the delay time distribution, and the changing 
scattering properties, as nuclei slide lower in charge, to arrive a 
predicted chemical composition and spectrum at Earth.

This type of ``principal component analysis" will have to be repeated 
for each source class, for which we have quantitative predictions using 
a complete sample, as above. It is to be expected that different sources 
have different chemical composition of their ultra high energy cosmic 
rays, and could appear as extremely different in such an analysis. Given 
sufficient statistics it might be possible to invert the procedure by 
assigning to each event a most probable source, and then adding up to 
obtain both the source spectrum, and the scattering distribution; this 
would have to be consistent with what we know about the source, its 
plausible chemical composition contribution, and magnetic fields. One 
obvious further consequence is that at very high energy the northern and 
southern sky should be different.

This will only be possible with an all-sky survey with matching 
sensitivity and observing procedures.

\section{Conclusion}

Gamma ray bursts are not yet ruled out, but would require very much 
higher fluxes from nearby sources such as M82 or NGC2146 and the like 
than expected based on gamma ray burst statistics (Pugliese et al. 
2000). The predictions are not sufficiently reliable to completely rule 
out or confirm such an idea. However, the cloud of events around Cen A 
would suggest in such a picture, that the starburst in Cen A actually 
produces a sufficiently large number of gamma ray bursts so as to 
dominate the sky distribution. This would be quite compatible with the 
Das et al. (2008) scattering simulations, and also the air fluorescence 
data from HiRes (Sokolsky 2008). If so, the air fluorescence data 
obtained within Auger should confirm a pure proton composition. On the 
other hand, if it is true, that a subset of Wolf-Rayet stars explode as 
gamma ray bursts, could it be that gamma ray bursts also pick up the 
abundances from the wind shells, as supernovae are believed to do? Such 
a picture would lead to a very similar high energy spectral behaviour of 
the different chemical elements; two problems, however, appear for this 
line of thinking: 1) Gamma ray bursts have a rapidly decreasing Lorentz 
factor with time, and so a final spectrum will be extremely smeared. 2) 
The Lorentz factor in gamma ray bursts is so high, of order 300, that 
acceleration from the knee region would go far beyond 30 EeV, and then 
there would be no spectral downturn, at least if the main sources are 
just 3 Mpc or so distant. This might deserve a dedicated test simulation.

Radio galaxies still provide the best bet to explain the data, but do 
face a number of serious difficulties. We have shown how to overcome 
such problems in the physics interpretation, and have suggested how to 
deal with the coming data. In a speculative approach we suggest strong 
substructure in the interstellar medium, and also the intergalactic 
medium, and also suggest the chemical composition spectral structure at 
very high cosmic ray energies in the context of a starburst (Biermann \& 
Fricke 1977) model: The chemical composition at the knee of galactic 
cosmic rays, derived from exploding Wolf-Rayet stars (Stanev, Biermann, 
\& Gaisser 1993) is reproduced at the highest energies, with just a 
factor-shift in energy and flux for all particles (Gallant \& Achterberg 
1999); this leads to a sequence in energy from lighter towards heavier 
nuclei, just as at the knee. We furthermore strongly predict, that given 
Cen A as the adopted source, the observed cosmic ray particle at high 
energy must be heavier nuclei, such as Carbon, Oxygen and heavier. We 
suggest a global strategy to deal with complexity of photo-dissociation, 
delay times, angular scattering, and range of possible sources 
detectable at Earth; the only way to overcome these difficulties is to 
use a $4 \pi$ sky survey with matching procedures and sensitivity.

In terms of ultra high energy cosmic rays radio galaxies come in two 
classes, those with an associated starburst with exploded Wolf-Rayets 
stars as feeding source such as Cen A, and those with just the 
inter-stellar/-galactic medium as a feeding source such as M87. Right 
now much of the known data suggest that Cen A could be the single 
dominant source. The two different classes of radio galaxies will look 
very different in arriving cosmic rays, and will also likely look 
different in TeV $\gamma$-emission and high energy neutrinos.

The future promises to be exciting in this field.

\section{Acknowledgements}

Collaborations and discussions with Eun-Joo Ahn, Rainer Beck, John Belz, 
Johannes Bl{\"u}mer, Silke Britzen, Alex Curu\c{t}iu, Ioana Du\c{t}an, 
Carlos Escobar, Tom W. Jones, Hyesung Kang, Marina Kaufman, Timo 
Kellmann, Gopal Krishna, Phil Kronberg, Sera Markoff, Gustavo 
Medina-Tanco, Joao de Mello-Neto, Faustin Munyaneza, Biman B. Nath, 
Giovanna Pugliese, J{\"o}rg P. Rachen, Gustavo Romero, Markus Roth, 
Dongsu Ryu, Vitor de Souza, and Arno Witzel are gratefully acknowledged. 
Work with PLB was supported by contract AUGER 05 CU 5PD 1/2 via DESY/BMB 
and by VIHKOS via FZ Karlsruhe; by Erasmus/Sokrates EU-contracts with 
the universities in Bucharest, Cluj-Napoca, Budapest, Szeged, Cracow, 
and Ljubljana; by the DFG, the DAAD and the Humboldt Foundation; and by 
research foundations in Korea, China, Australia, India and Brazil. 
Support for JKB comes from the DFG grant BE-3714/3-1 and from the 
IceCube grant BMBF (05CIPE1/0). This research has made use of the 
NASA/IPAC Extragalactic Database (NED) which is operated by the Jet 
Propulsion Laboratory, California Institute of Technology, under 
contract with the National Aeronautics and Space Administration. This 
research also made use of the ViZier system at the Centre de Donne{\'e}s 
astronomiques de Strasbourg (CDS) (Ochsenbein et al. 2000).

\end{document}